# New indicators for assessing the quality of *in silico* produced biomolecules: the case study of the aptamer-Angiopoietin-2 complex

R.Cataldo[a], L. Giotta[b], M. R. Guascito[b], E.Alfinito[c,]*

[a.] *Department of Mathematics and Physics, "Ennio de Giorgi", University of Salento, Via Monteroni, Lecce, Italy, I-73100*
[b.] *Department of Biological and Environmental Sciences and Technologies, University of Salento, Via Monteroni, Lecce, Italy, I-73100*
[c.] *Department of Innovation Engineering, University of Salento, Via Monteroni, Lecce, Italy, I-73100*

*Corresponding Author: eleonora.alfinito@unisalento.it

**Abstract**

Computational procedures to foresee the 3D structure of aptamers are in continuous progress. They constitute a crucial input to research, mainly when the crystallographic counterpart of the structures *in silico* produced is not present. At now, many codes are able to perform structure and binding prediction, although their ability in scoring the results remains rather weak. In this paper, we propose a novel procedure to complement the ranking outcomes of free docking code, by applying it to a set of anti-angiopoietin aptamers, whose performances are known. We rank the *in silico* produced configurations, adopting a maximum likelihood estimate, based on their topological and electrical properties. From the analysis, two principal kinds of conformers are identified, whose ability to mimick the binding features of the natural receptor is discussed. The procedure is easily generalizable to many biological biomolecules, useful for increasing chances of success in designing high-specificity biosensors (aptasensors).

**Introduction**

The development of new methods in medicine and pharmacology involves the use of very specific and targeted macromolecules, useful in diagnosis and therapy. In this framework, small fragments of RNA and DNA, called aptamers, are arising high interest. Since 1990, the SELEX technique [1] is used to artificially generate high affinity aptamers able to bind a specific protein/molecule. SELEX is potentially able to produce, from an initial very large pool of oligomers, those with the highest affinity for the specific target; however, it is labor-intensive, time-consuming and requires high resources and cost of finances [2]. This technology is an iterative process that foresees essentially two phases, selection and amplification [3][and references therein], followed by amplification and conditioning [3].

At present, an optimization of the SELEX technique (Cell-SELEX) has been developed, leading to selected structures able to bind a wider range of targets, from viruses to live cells [3, 4], becoming a reference tool for selecting aptamers of interest in medicine and biotechnology [5, 6].

Despite the huge amount of works reporting on the selection of new aptamers recognizing specific biological targets, the current understanding of chemical-physical mechanisms responsible for their binding specificity is limited. Gelinas et al. [7] suggested that, in most cases, aptamer binding occurs without a conformational change of the target, while either population-shift [8] or induced fit [9] mechanisms have been considered for describing the ability of aptamer ligands to bind specifically the relevant targets. Wlodarski and Zagrovic [10] highlighted the role that the timescale of conformational transitions plays in controlling binding mechanisms. Recently Autiero et al. [11], based on their molecular dynamics simulations, pointed out that intrinsic flexibility of aptamers may be essential for partner recognition. Furthermore, a number of experimental methods for measuring aptamer-protein binding equilibria have been employed and recently reviewed by Jing and Bowser [12].

In parallel with *in vitro* procedures for selection and evaluation of aptamer sequences, *in silico* approaches have been developed to predict the 3D structure of high-affinity aptamers. *In silico* predictions are of great help in the identification of aptamers with high affinity for the specific target, reducing time and cost in the screening of the aptamer families [13]. However, despite the advancements made in the last few years [14] [and references therein], lack of information still persists and prevents an extensive comprehension of binding functionalities. Among the computational approaches, docking procedures are able to predict the ligand conformation, when the binding pocket of the target is given, and to score the results on the basis of the foreseen biological activity [15]. However, scoring still remains a quite complex task, due to a plethora of different variables that intervene in the real compound formation, as well as the change of conformation and/or the role of solution [16]. The transition between different conformers, in particular, is a quite relevant process that presides at the formation of the receptor-ligand complex. The complete dynamical evolution behind the formation of the complex is still an open problem, mainly classified into the two above mentioned scenarios: the population-shift, also known as conformational selection, i.e. the selection of the binding-competent structure due to its specific conformation (the conformational change happens before the binding); or the induced fit, i.e. the subsequent adaptation of a loosely bound conformer to the active site of the receptor (the conformational change happens after the binding) [17]. Most of the common *in silico* procedures implement a rigid docking [18], although other methods have been developed to incorporate the induced fit mechanism [19, 20].

In the present paper, we consider a set of 4 anti-Angiopoietin-2 (Ang2) aptamers and one anti-Angiopoietin-1 (Ang1) aptamer (Sequence 16), used as a control, initially studied by Hu and co-workers [21]. Angiopoietins (specifically Ang1 and Ang2) have a prominent role in vasculature upregulation in many kinds of tumors, thus making these proteins of high interest in therapy and early detection [22]. Specifically, the Ang1- Ang2 interaction is not completely clarified and their monitoring is of high relevance also for avoiding sepsis shocks in patients treated with chemotherapy [23]. The authors of [21] selected RNA mutant sequences with high affinity for Ang2, starting from the sequences of anti Ang2 aptamers, obtained by *in vitro* procedures. By using the ZDOCK program, Hu et al. [21] scored the aptamer-protein interactions of each sequence. Then experimental data were obtained from measurements with a surface plasmon resonance (SPR) biosensor, to test the prediction accuracy. The three highest score mutant sequences, along with a high and low affinity binding sequence were analysed [21]. The paper highlighted that one of the mutant aptamers (Sequence 2_12_35) was one of the worst performing in experiments [21], although it showed the best computational rank among the five considered aptamers. Hu et al. [21] pointed out the conflicting results concerning performances expected for the *in silico* produced aptamers and observed in the *in vitro* produced aptamers.

In a recent paper [24], we outlined a computational strategy, based on free software, i.e. SimRNA and AutoDock-Vina [18, 25] to produce a set of possible configurations for the 5 aptamers proposed in [21]. This strategy produces conformers whose expected affinity is in line with the measured one [21].

Here we are interested in going deeper inside, focusing on a strategy able to select the configurations with the highest probability to be found *in vitro*. In doing so, we will follow the pathway of Proteotronics, a procedure for conjugating structure and function of proteins and aptamers at a microscopic level[26]. Structure and function of biomolecules can be described simultaneously, by using a complex network whose topology pictures the backbone arrangement; the network represents a set of selected interactions among the biomolecule elements. Interaction may depend on the environment, specifically on the presence of appropriate targets[27-31]. The centre of our attention is now on the electrical interaction in a linear regime that allows the investigation of the network at a local level, by means of the percolation of electrical charges. In particular, we perform a statistical analysis of some topological and electrical features of the numerous configurations produced by the strategy outlined in [24], introducing some quality indicators that will be in turn tested against the experimental data. In doing so, we outline a procedure to select the configurations with the highest probability to be found in *in vitro* or *in vivo* experiments.

As a general result, we find marked differences among the structures of Ang2 and Ang1 specific sequences, in the whole amount of produced configurations; for that reason, we can roughly classify them into two large families, really not very different with respect to the calculated affinity. This could mean that part of these configurations has low effectiveness in the agonistic/antagonistic mechanism and is finally cut off, or that it can evolve toward the most effective configuration. In other terms, without the claim of a new proposal for solving the well-known dilemma between conformational selection and induced fit [32], we note that, at least at the theoretical level (ranking), the quality of aptamer-ligand interaction plays a crucial role in the conformational agreement, which could not be evaluated only ranking the poses.

## Materials and Methods

We deal with the same problem as in [21], i.e. a comparative evaluation of the binding to Ang2 of five different aptamers, specifically:

1. an aptamer, denoted "Sequence 1", both in Hu's paper and below, from the pool of Ang2 specific RNA aptamers known in the literature;

2. three mutant sequences, here and in Hu's paper denoted "Sequence 2_12_35", "Sequence 15_12_35", and "Sequence 15_15_38";

3. an Ang1-specific RNA aptamer, denoted "Sequence 16", as in [21], there applied as a control sample.

The primary structure of the aptamers is given in [21], while the 3D structure of the target protein [33] is reported in the public databank [34], as a fraction of the complete Ang2 protein (from E280 to D495, chain A, entry 1Z3S), representing the receptor binding domain.

The procedure of structure prediction is detailed in [24]; here we recall that the 3D structures of the aptamers were sampled by using the SimRNA tool, a stochastic procedure based on the Replica Exchange Monte Carlo simulations [25]. Then, the most stable, i.e. the lowest energy, configurations are selected from the huge number of possible structures, and docked to the receptor, by means of the AutoDock-Vina tool [18]. Both SimRNA and AutoDock-Vina are free packages.

From hereafter, we will call SiRVA the cascade procedure SimRNA + Autodock-Vina, summarized into two steps as follows:

a. SimRNA simulation produces a set of stable configurations [25], starting from the nucleotide sequence of the single aptamer. These configurations are ranked on the basis of their SimRNA potential value [25]; this dimensionless quantity, say $\varepsilon$, works like a thermodynamic potential, therefore, it is reasonable to represent it as: $\varepsilon = E_{SimRNA}/RT$ where R is the gas constant and T is the room temperature [24].
b. The docking procedure is performed by using AutoDock-Vina [18]. To this aim, the most stable SimRNA configurations were tested for docking with the Ang2 fragment 1Z3S [33]. MGLTools [35] was employed to assign partial charges and atom types to the aptamers [24]. It roughly consists in a rigid docking of the ligand by using 9 different space rotations for each configuration, each of them characterized by an energy value, $E_{dock}$, also called *affinity* by AutoDock-Vina. The first element of each cluster has the best score; the number of clusters strongly depends on the sequence, with significant differences (Table 1). The unsuccessful poses were excluded by visual inspection or evaluation about the $E_{dock}$ values, giving a total number of reliable realizations produced by the SiRVA, reported in Table 1. We rank the docked configurations through a new indicator, called Effective Affinity (EA), calculated as the total energy, EA= $E_{dock} + E_{SimRNA}$ [24].

**Table 1.** Number of reliable realizations produced by the SiRVA procedure [24]: the number of clusters for each sequence, the number of docked structures.

| Sequence | 1 | 15_15_38 | 15_12_35 | 2_12_35 | 16 |
|---|---|---|---|---|---|
| # of survived clusters | 20 | 11 | 21 | 38 | 27 |
| # of survived docked configurations | 178 | 89 | 156 | 321 | 236 |

At present, the selection of the best performing configuration, or set of configurations, is not automatic, since data from both procedures (SimRNA and AutoDock-Vina) have to be considered.

**The Proteotronics approach** The *Proteotronics* approach is a theoretical procedure able to analyse the physical response of biomaterials in electronic devices[24, 28-31, 36-38]. Proteotronics models the single macromolecule at the microscopic level, combining information from both structure and function. The macroscopic physical features emerge from the microscopic interactions and reflect the 3D structure. Different procedures are available in the literature (see, for example, refs. [39-41]), whose level of refinement ranges from the complete molecule to the single atom. Proteotronics works at the single amino-acid (nucleobase) level, which is sufficient to keep most of information useful for technological applications, with the advantage of small computational time. This model has been recently extended to aptamers[38], with encouraging results.

The procedure consists of three steps:

- The graph analogue building;
- The interaction network building;
- The network solution.

The graph analogue building starts from the 3D structure of the biomolecule, i.e. its space organization at the atomic level, recorded in specific experimental conditions [34]. We select the carbon atoms $C_1$ ($C_\alpha$) to represent the position of each nucleobase (amino acid). Then, the neighbor nodes are connected with a link and the adjacency matrix can be written down[28]. The neighboring statement depends on the value of a free parameter, the cut-off radius $R_C$; two nodes are defined neighbors when closer than $R_C$. In such a way, we have the possibility to explore different possible graphs, each of them corresponding to the same macromolecule, at different levels of activation[29, 31, 42].

The graph is then converted into an interaction network, in the present case, an electrical network to represent the flowing of charges inside the macromolecule. A resistance is assigned to each link. In particular, the resistance between a couple of nodes, say *a,b*, is calculated as that of a cylindrical structure of length $l_{ab}$, the distance between the nodes, and surface $A_{ab}$, the intersection area of the spheres of radius $R_C$, drawn around the nodes[26].

About the network solution, the physical response is calculated by assuming a couple of ideal electrodes connecting the network to an external bias. The network is then solved, for an assigned value of $R_C$, by using the standard Kirchhoff's laws. The output is the network resistance, calculated in the linear regime, by using appropriate resistivity values, as detailed in[29]. The global resistance depends not only on the number of links, but also on the kind of pathways the charge follows, i.e. its diffusion conveys the presence of bottlenecks, dead ends, and so on.

The resistance monitoring has revealed an effective method not only to explore the global topological properties of the structure, but also the spreading of an epidemic on a complex network, giving information on the network itself [43].

## Results

**Link number vs. Effective Affinity.** A signal of high binding affinity is a good structure complementary between the receptor and the ligand [24]. In terms of network, this means that a significant growth of the link number has to be observed, when the receptor is added to the ligand. Therefore, it is interesting to explore to what extent the growth of the link number and the effective affinity are related. In such a way, it is possible to assess whether EA is a good descriptor of the structure topological properties. In Figure 1, EA is reported vs. the relative link number, i.e. the difference between the link number of the ligand-receptor complex and the link number of the aptamer alone. Each point corresponds to a single configuration.

The correlation between EA and the number of links (see also the Spearman correlation in Table 2) says that EA is a good indicator, at least for what concerns the topological content of the binding-affinity concept.

**Resistance response.** Another kind of investigation concerns the resistance of the electrical network corresponding to an assigned structure (the free-target aptamer and the aptamer docked with the protein).

The global resistance of the network is calculated for different values of $R_C$, to obtain a resistance spectrum. It reflects the global network connectivity and the link distribution in the space. For very large $R_C$ values, the resistance tends to an asymptotic value, sequence-specific. Figure 2 reports the ratio between the resistance of the complex and the resistance of the target-free aptamer ($r_{comp}/r_{apt}$). Each curve represents the average over all the configurations of each sequence. We can notice that, by adding the target protein, the resistance becomes smaller than the resistance of the aptamer alone. This trait is common to other kind of complexes[29] and says that the protein well fits the aptamer, efficaciously completing its network with many parallel-resistance links.

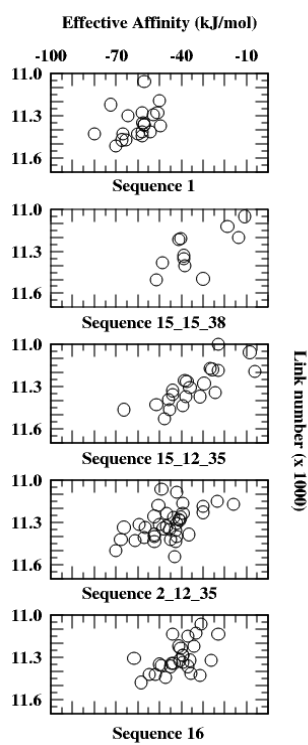

**Figure 1.** The relative link number vs. the effective affinity (EA) for the 5 structures. Each circle corresponds to a single realization.

**Table 2**. The Spearman correlation between EA and the relative link number.

| Sequence | Rank | Significance |
|---|---|---|
| 1 | -0.50 | 2.93 10$^{-2}$ |
| 15_15_38 | -0.58 | 6.58 10$^{-2}$ |
| 15_12_35 | -0.83 | 2.06 10$^{-4}$ |
| 2_12_35 | -0.57 | 4.75 10$^{-4}$ |
| 16 | -0.48 | 1.49 10$^{-2}$ |

From Figure 2, it is evident that Sequence 16, the Ang1-specific aptamer, has the smallest relative resistance in the whole $R_C$ range. The relative resistances of the Ang2-specific sequences are quite close each other, all over the spectrum. Among them, Sequence 2_12_35 has the lowest curve all over the range. We can notice that on a wide range (until about 50 Å), the relative resistances of all the sequences are sorted in ascending order of the binding constant, $K_a$, given in [21], and that the smaller $R_C$, the marked the differences among structures. As a matter of fact, at small $R_C$ values, only the links necessary to obtain a stable configuration are present, while increasing $R_C$, many not necessary links may appear [44]. However, the possibility to discriminate Ang1 from Ang2-specific aptamers persists also in the asymptotic limit. In agreement with previous results[27, 29], we conclude that resistance is an interesting tool for affinity investigation, especially in the linear regime.

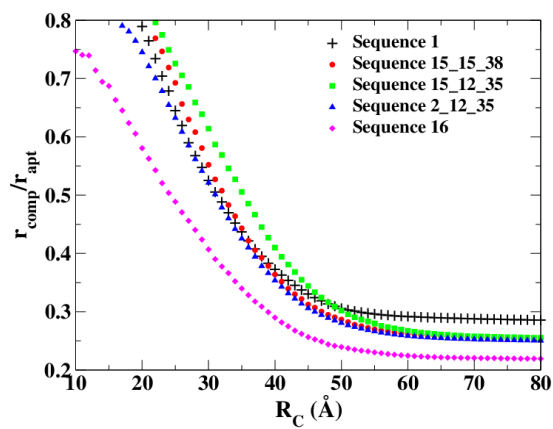

**Figure 2**. The relative resistance spectrum of the 5 sequence. Each curve results as the mean value of all the realizations of the selected sequence.

In the following, we will try to identify some indicators useful for selecting, for each aptamer, the set of complexes that have more chances to be found *in vitro*.
First, we notice (see Figure 1) that the energy landscape explored by the SiRVA procedure is quite frustrated, with the presence of few outliers. This suggests that, beyond energy/effective affinity, other information is necessary to rank the configurations, for example, data relative to their topology. We expect that the computational procedure symmetrically converges *in structure and function* (best 3D structure and best chemical affinity) toward the real topology of the complex. Therefore, we postulate that the pose which best describes the real complex has to be that which optimizes not a single but a set of different descriptors.

**Definition of reliability indicators.** Following the idea that SiRVA procedure converges toward the most probable configurations, we aim at identifying the main features of these configurations. Therefore, we focus on their topological properties, since a good topological fit agrees with a good energy fit (see Figure 1). There is not a unique way to select the relevant topological features of the aptamer-protein complex; information comes from both local and global properties. Therefore, we look for indicators concerning both these aspects; as a local information, we take the link number of each node of an assigned configuration, and as global information, the set of nodes to which each node is connected. The mean value of these indicators, calculated as reported in the Appendix and called M, is used, together with the effective affinity EA, to estimate the configurations, which have the highest probability to be found as real products.

In Figure 3, we plot these indicators on single bar-graph; EA values are distributed between 1 and $N_{seq}$ as in Eq. (A3), and called $EA_{ord}$, $EA_{ord}=1$ indicates the best score, $EA_{ord}=N_{seq}$, the worst. Finally, to make the comparison among the sequences easier, the values of both $EA_{ord}$ and M were normalized to 1. Hereafter, a plot like that of Figure 3 will be called *shadow plot* (SP), because one of the quantities to be represented, here $EA_{ord}$, is drawn with the opposite sign, like the shadow of the topological indicator.

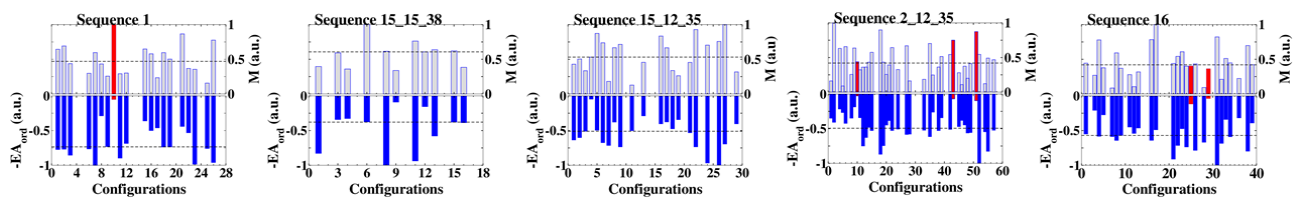

**Figure 3**. Slide show of the shadow plots of the 5 sequences. Dashed lines correspond to medians, outliers are red coloured.

In many cases, the lowest $EA_{ord}$ value is an outlier (bottom outlier) in the Tukey definition [45] (more the 1.5 times the *interquartile range* (IQR), or approximately 3 standard deviations in a Gaussian distribution), and this makes its reliability doubtful. In general, we pay careful attention to outliers, because they can signal skewed distributions instead of a statistical error in a unimodal and symmetric distribution [42]. However, in the present case, the numerical strategy used for producing the configurations does not open the door to this kind of scenario. Therefore, the configurations showing anomalous values of $EA_{ord}$ (in Sequences 1, 2_12_35 and 16) will be considered not interesting. In particular, configuration 10 of Sequence 1 shows anomalies in both M and $EA_{ord}$; configurations 10, 43 and 51 of Sequence 2_12_35 are anomalous in $EA_{ord}$; configurations 25 and 29 of Sequence 16 are anomalous in EA. Neither Sequence 15_12_35 nor Sequence 15_15_38 have bottom outliers in EA$_{ord}$ or in M. By inspecting the SPs, we can observe that the outliers in EA$_{ord}$ correspond to extreme values in M, with the noticeable exception of Sequence 16. Finally, there are many structures with low values of both EA$_{ord}$ and M: they are the natural candidates to represent the real complexes.

The boxplots in Figure 4 schematically resume the distribution of M and EA$_{ord}$ for all the sequences. Both these quantities are normalized to 1. Outliers of M are present only in the right side of the plot, above the forth quartile, while those of EA$_{ord}$ can be found principally in the left side of the plot, below the first quartile.

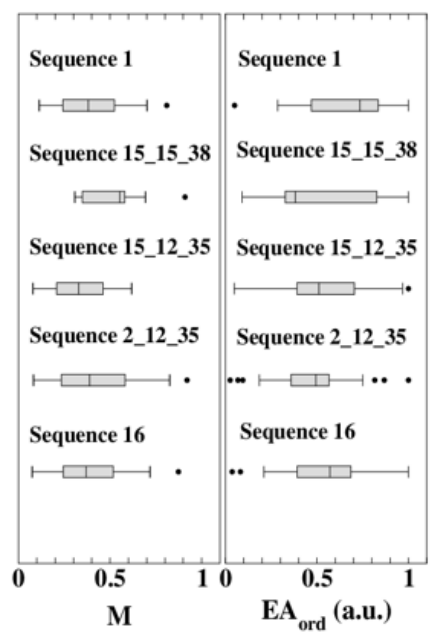

**Figure 4.** Boxplots of the proteotronics indicator M and the chemical indicator EA$_{ord}$. Both quantities were distributed between 0 and 1.

Concerning the extreme values, in Sequence 2_12_35, the percentage of EA$_{ord}$ outliers is about 16%, much larger than the other sequences (less than 6%). At present, we are not able to conclude whether the pose generation of this sequence failed in some critical points or , for example, whether the proposed primary sequence is not appropriate for binding Ang2 with high affinity, however, as previously shown and in agreement with other information in the literature, Sequence 2_12_35 has low affinity with Ang2.

**Scoring and Conformers.** Finally, we compare the features of the computational produced structures with those of the Ang2-Tie2 complex. The full-length Angiopoietin-2 protein is composed of a single amino acid chain folded in three main domains[46]: the N-terminal super-clustering domain (SCD), a central coiled-coil domain (CCD), responsible for Ang2 dimerization, and a C-terminal fibrinogen-related domain.

which represents the receptor binding domain (RBD). Three main subdomains (A, B and P) can be identified in RBD, being P the outermost one. Crystallographic data show that the Ang2 RBD is bound to the Tie2 Ig2 domain at the level of the P-subdomain, where the Lys 469, Lys 473 and Tyr476 residues, essential for binding [33], are located. It is believed that in the Ang2 dimeric or multimeric forms, the P subdomains are oriented away from the dimerization interface, allowing the simultaneous interaction with two distinct Tie2 receptors. [33]

Our computational procedure for the selection of aptamer conformers in the Ang2-bound state has been carried out using the available 3D structure of the Ang2 RBD domain, whose surface has been theoretically considered fully accessible for binding. However, we expect that *in silico* generated complexes reproduce at least partially the kind of binding established between Ang2 and its natural receptor. It has been shown that Ang2 P subdomain binds at the tip of the Tie2 Ig2 arrowhead through a lock-and-key mechanism, where two complementary surfaces interact with each other with no domain rearrangements and little conformational change in both counterparts.[47] In general, we find that our aptamer-Ang2 complexes mainly exhibit two kinds of binding-competent conformers: the aptamer located on the head of the P subdomain, hereafter called *hair*, or the aptamer embracing the protein, thus interacting also with A and B subdomains, hereafter mentioned as *belt*. Although the hair configuration is more similar to that of the Ang2-Tie complex than the belt configuration (see Figure 5), a marked difference in the EA values is not observed. A belt configuration may be characterized by a high value of EA, so as the hair configuration, it happens when the aptamer is very close to the protein, i.e. it contains many links (see Figure 1). In Figure 5 the cartoon of the Tie2-Ang2 complex is reported with the corresponding contact map, calculated for $R_C$=20Å.

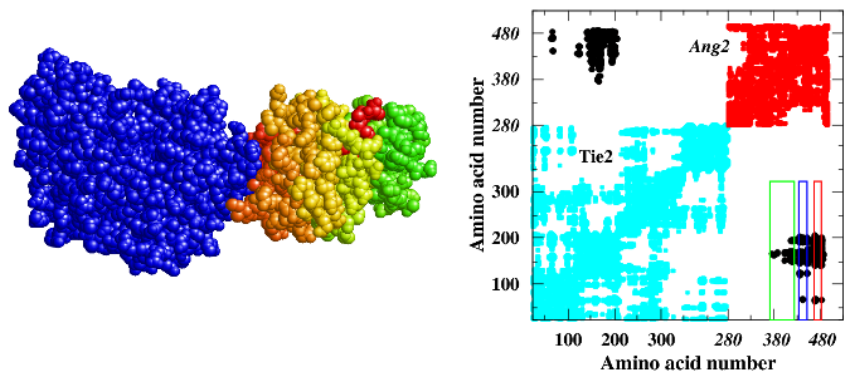

**Figure 5.** Cartoon of the Ang2-Tie2 complex (left), with Tie2 in blue, and the corresponding contact map for $R_C$=20Å (right). In the contact map, links internal to Ang2 fragment, from E280 to D495 are drawn in red; links internal to Tie2 fragment, from A23 to P445 are drawn in cyan. Links between the two proteins are drawn in black.

In Figure 6A, a slide show of the most representative complexes of the 5 studied sequences and the corresponding contact maps is given. In each contact map, we have selected three regions of possible binding with decreasing level of connectivity: the first (in red) from N467 to Y482, the second (in blue), from I434 to S453, the third (in green), from V370 to G420. These regions belong to the P subdomain according to the features of the natural Ang2-receptor complex. Accordingly, a hair conformer has the largest part of points (i.e. network links) in these three regions; otherwise, belt configurations have preferential attachment away from the P subdomain.

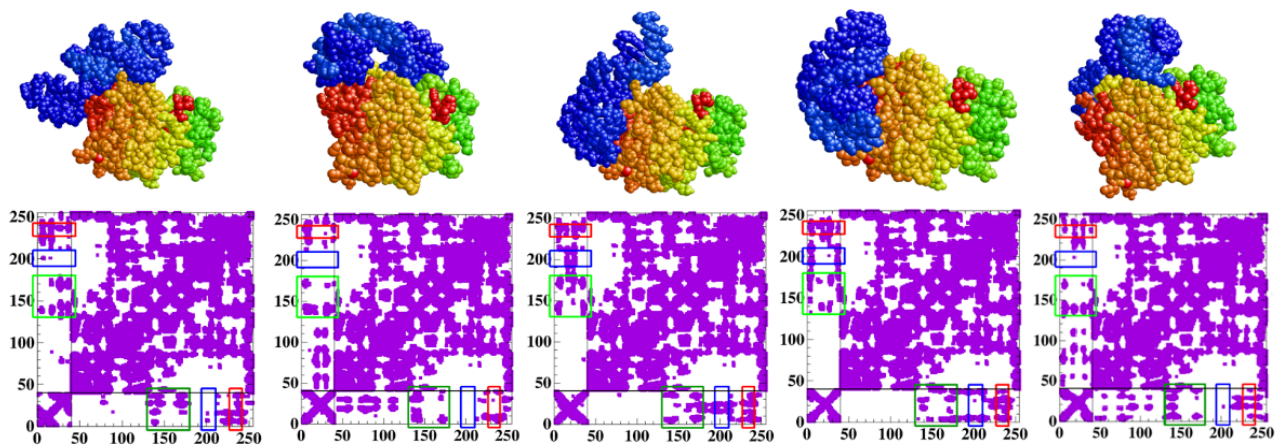

**Figure 6.** Slide show of the most representative complex of the 5 studied sequences (top line) and of the corresponding contact maps (bottom line) at $R_C$=20Å. From left to right: Sequence 1(8), Sequence 15_15_38 (9), Sequence 15_12_35 (4), Sequence 2_12_35 (36) and Sequence 16(3). The first 40 (41 for Sequences 15_*_*) nodes of each map compete to the aptamer, following nodes compete to the protein fragment, from E280 to D495. The boxes correspond to the three region of binding in the Tie2-Ang2 complex (see text).

As outcomes of our computational experiment, we observe that the frequency of belt/hair configurations depends on the particular sequence; therefore, we will give a paramount of those findings, before attempting a discussion on possible selection criteria.

Considering the EA and M values in Figure 3 and in Supplementary Information (SI), we can notice that:

- In Sequence 1, the highest values of EA pertain to realizations 10 (belt), which is an outlier, and 8 (hair). Configuration 8 is acceptable, since its M value is quite small (lower than the median). Configurations 15 and 21, which have good values of EA are, otherwise, too far from the average topological features, so they are not selected; therefore, the second choice is the hair configuration 17 with quite good values of EA and M.
- In Sequence 15_15_38, the highest EA values pertain to configurations 9 (belt) and 12 (belt). Configuration 9 is below the median of M; this is not true for configuration 12. Going to the 3th highest value of EA, configuration 4(hair) is below the M median and can be taken as the second choice.
- In Sequence 15_12_35, the highest EA values pertain to configurations 4 (hair) and 13 (belt); both have M below the median. The third value of EA pertains to 19(belt) that also has M below the median.
- In Sequence 2_12_35, the highest EA values pertain to configurations 10, 43, 51 (belt) which are outliers. Configurations 9 (hair), 4 (hair), 36 (hair) and 5 (belt) follow in EA values. Configurations 9 and 4 are too far from average topological features, thus only 36 and 5 can be chosen.
- In Sequence 16, the best EA values pertain to realizations 24 (hair), and 29 (belt), both these configurations are outliers, and have very few links in the original binding pocket. Going to the 3$^{th}$ best EA value, configuration 3 (belt) has M below the median and can be chosen as the most representative. The 4$^{th}$ best EA value pertains to configuration 5 (belt), which has M below the median and can be chosen as the second choice.

In order to optimize M or EA, thus cutting off a number of configurations, different criteria could be adopted, like, for instance, picking the poses that are in the first quartile both in M and in EA; the list of these configurations is reported in SI.

In most cases, the best value of EA has also a good value of M, confirming the hypothesis that the computational procedure converges toward the most probable configurations. Good values of EA do not discriminate between hair and belt conformers, indicating that they would be both probable products of the binding reaction in a population-shift scenario. In particular, by selecting only the pose that optimizes both M and EA, we obtain a belt configuration for the Ang1-specific aptamer, and a hair configuration for the Ang2-specific aptamer instead [48] (see SI). Interestingly, the mutant sequences have both hair and belt configurations.

Widening the perspective of possible events, we can imagine that in presence of the isolated RBD domain the binding reaction product may appear *in vitro* in more than one conformer as a consequence of enhanced conformational flexibility and subsequent adaptability of the aptamer compared to the Tie2 receptor, thus paving the way to an unconventional (bifurcated) population-shift scenario. On the other side, an induced fit scenario involving a post-binding transition from the belt to the hair configuration can be ruled out, since the belt conformer in most cases is not "loosely-bound" with respect to the hair one, being both tightly bound. Moreover, belt and hair conformers are structurally too different to imagine a facile conversion from belt to hair with a sufficiently low activation barrier.

In a population shift perspective, the computational selection of only hair poses could be considered an indication of high specificity (i.e. enhanced receptor mimicking ability) of the considered aptamer sequence, whereas do not select a stable hair conformer would represent an indication of low specificity. From this point of view, our combined energetic and topological approach is extremely successful in discriminating the best (Ang2-specific aptamer) from the worst (Ang1-specific aptamer) sequences able to bind Ang2 protein.

When crossing our data with experimental affinity tests carried out by Hu et al [21], some important issues must be considered: their SPR sensor was constructed by immobilizing the biotinylated aptamers on the metal surface exploiting the thiol-gold covalent binding and the biotin/streptavidin linkage strategy. This means that the Angiopoietin-2, flowed onto the functionalized sensor surface, was free to interact with the aptamer virtually in any orientation. However, it is noteworthy to point out that Hu and coworkers did not employ as ligand the Ang2 RBD domain, but the commercial full-length protein, which of course is expected to dimerize, through the CCD, and/or to form higher order multimers, whose formation is facilitated by the N-terminal SCD. Since it has been suggested that A and B subdomains in the RBD moiety could mediate interactions between the individual angiopoietin monomers in the multimeric assemblies [47], it is possible to assume that these RBD subdomains are partially or totally inaccessible for aptamer binding. In the light of these considerations, the SPR experiment does not allow to test the actual affinity of most belt conformers toward the RBD domain, as these conformers embrace RBD at the level of A and B subdomains.

This finding nicely explains the noteworthy correspondence of our prediction results with experimental data, if some belt configurations are rejected based on supramolecular steric hindrance.

## Conclusions

This paper analyses the pool of *in silico* generated 3D configurations of an assigned biomolecules and proposes a procedure to identify those having the largest chance to represent the real structure. Investigation is performed by using a complex network approach, which maps the biomolecule into an electrical network, both the target-free aptamer and the aptamer-protein complex. In such a manner, the topological properties of the biomolecule are preserved together with its resistance response, when contact with an external bias is simulated. Therefore, both local and global topological features of the biomolecule can be explored, especially binding affinity, in terms of the ability to bind the cognate molecule, and stability of the binding. The first item is linked to the structure of the complementary biomolecules (receptor and ligand), technically described by the association constant, $k_a$, while stability is described by the binding affinity constant, $K_a$.

The procedure is applied to a set of 5 anti-Angiopoietin aptamers: four with demonstrated high affinity for Ang2, and one, Ang1- specific, used as a control; their 3D structures were produced *in silico* as in [24], for the oligomer alone and complexed with the protein. The present strategy introduces both physical (M) and chemical (EA) indicators to select configurations with the best affinity and the highest probability to be found in experiments. To test the accuracy of the selection produced by this couple of indicators, we check the presence of the expected links between the aptamer and some specific amino acids of Ang2. A rough, although efficacious, classification selects the *in silico* produced configurations into 2 types : hair-like, which is the most similar to the original Ang2-Tie2 binding, hanging the aptamer on the outermost P subdomain of the Ang2 RBD, and the belt-like, in which the aptamer embraces the protein with a small binding with the original site and supplementary contact points with A and B subdomains. Not relevant differences in the EA indicator were found between the hair and belt configuration, although the frequency with which they appear in the different aptamers, strongly suggests they are related to a low stability kind of binding. A conclusive world could be only said by a crystallographic analysis.

As a general result of this investigation, the difficulty in making a reliable ranking of in *silico* produced structures is even more emphasized; more elements have to be considered in performing the computational selection, such as, like in the present case, the role of dimerization in *in vivo* working biomolecules. This can reduce the surface on which the target may really be bound; due to the extreme flexibility of the aptamer, also in the case of a good/sufficient binding in the appropriate binding domain, parts of its body may end up on a not allowed surface of the receptor. Therefore, in the present state of the art, alternate investigations are necessary to complement and interpret the results of computational structure modelling.

The procedure here shown is simple to implement, easily generalizable to many biological biomolecules and can be considered as a complementary test to the ranking performed by standard computational approaches, useful for increasing chances of success in designing high-specificity biosensors (aptasensors). Moreover, the results are in good agreement with the present literature [21] and are valuable in the absence of crystallographic data.

## Appendix

In this appendix we detail the procedure for calculating the topological indicator M. As discussed in Section **Definition of reliability indicators**, we combine information from local and global features of the available configurations.

Concerning local information, for an assigned $R_C$ value (in present case, $R_C$ =20Å [24], we focus on the connectivity of each node, and compare it with its mean value, obtained performing the average all over the configurations. Therefore, we introduce an index, $I_1$, which is calculated as follows:

$$I_1 = \sum_{i=n}^{N} |x_i - <x_i>| \qquad (A1)$$

where: $x_i = \sum_{i<j} l_{i,j}$ , with $l_{i,j}$ =0,1, the link number between the nodes $i,j$; $<x_i>$ the same quantity averaged over all the configurations, $N$ is the number of nodes, n=1.This local information is combined with a global information, given by resistance. Resistance has to be meant as the response to the flowing of free charges inside a specific network arrangement, and monitors the global connection degree. For example, the presence of bottlenecks reduces the global resistance, even if the total number of links is high. The mean resistance spectrum of each sequence is then compared with the resistance spectrum of each configuration of that sequence.

A second indicator, $I_2$, is proposed, by using Equation (A1). In this case, $x_i$ is the resistance value of a configuration of a selected aptamer, calculated for an assigned $R_C$ value; $<x_i>$ represents the same quantity averaged over all the configurations of the aptamer. The range of $R_C$ values chosen for our investigation is between 10-110 Å, as reported in Figure 2, where the point $i$ of each curve corresponds to $<x_i>$. At the same time, for each structure, we calculate $I_3$ as the deviation from the average of the effective affinity (EA):

$$I_3 = |EA - <EA>| \qquad (A2)$$

These three indicators are finally distributed from 1 to $N_{seq}$ (i.e., the number of structures for an assigned sequence, see Table 1) by using the following transformation:

$$Y_k = 1 + (I_k - min)/\Delta \qquad (A3)$$

with: $\Delta = (max-min)/(N_{seq}-1)$ and $min/max$ the min/max value of $I_k$.

This transformation has two advantages: it uses the same range for all the indicators of a certain sequence, and preserves the distance between couples of Y-values, differently from a simple ordering. The mean value of $Y_1$ and $Y_2$, $M = (Y_1+Y_2)/2$ is also evaluated.

As a criterion for selecting the structures with the highest probability to be found *in vitro*, we propose to assume the closeness to the average of both topological and chemical indicators.

To identify the best performing topological indicator, we look for that which optimizes the correlation with $Y_3$. Therefore, for each couple of topological indicators ($Y_1$, $Y_2$ or M) and $Y_3$ we can classify the configurations in LL (both quantities below the respective medians); HH (both quantities above the respective medians); and LH (one quantity below and the other above the median). The choice of median value, instead of the average, helps to prevent any misinterpretation, due to the presence of several extreme values in the indicators.

In Figure A1, we plot these indicators on single bar-graph, hereafter called *shadow plot* (SP), in which $Y_3$ is drawn with the opposite sign, like the shadow of the topological indicators. The draw regards Sequence 1, as a pattern of any number of similar situation. The sets of LL, LH and HH configurations of each sequence are reported as pie-charts in Figure A2.

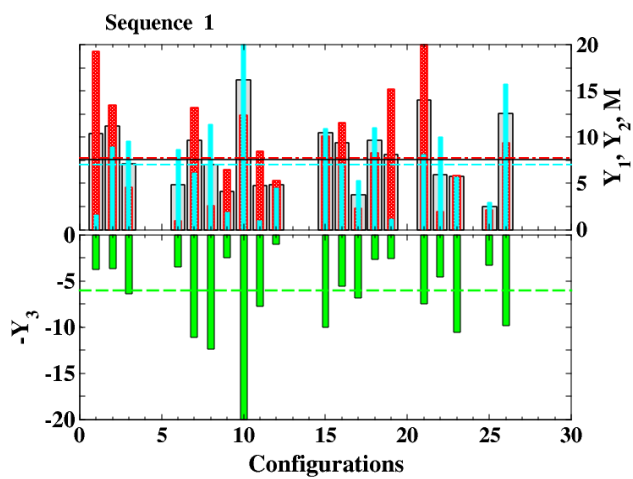

**Figure A1**. Shadow plot of Sequence 1. Lines indicate the median values of $-Y_3$ (green), M (black), $Y_1$ (red), $Y_2$ (cyan). The Y indicators as well as M are represented by bars: green for $-Y_3$, black for M, red for $Y_1$ and cyan for $Y_2$.

The LL and HH configurations show a convergence between the chemical and topological indicators (Figure A1). However, some not conclusive cases are present, in which the behavior of topological and chemical indicators is opposite (LH). Looking at the cumulative indicator M, we observe that the fraction of indeterminate cases (LH) is always not larger than the sum of the LL and HH cases, i.e. M optimizes the correlation with Y3 with respect Y1 and Y2, and gives more complete information about the configuration. Therefore, we can select M as the most useful indicator for searching, among the structures with very high effective affinity, those that best reproduce the expected characteristics of the real aptamer-protein complex. In other terms, we conjecture that a very high value of EA could correspond to a too extreme, not stable, topology, therefore, looking for a stable and reliable configuration, we have to maximize both the structural and chemical features.

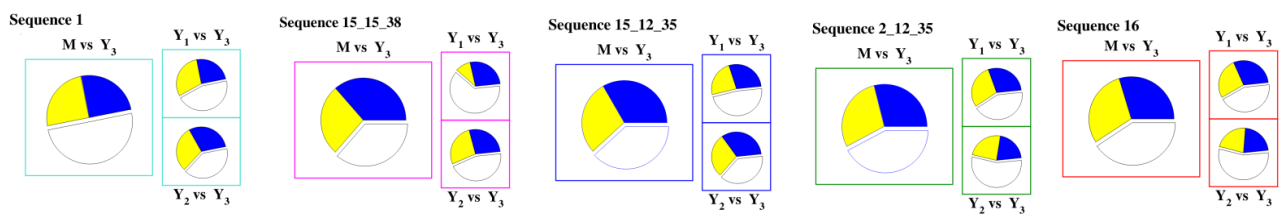

**Figure A2.** The pie charts of LL (blue), HH (yellow) and LH (white) realizations for the 5 selected structures. M (global index) data are compared with $Y_1$ (topological index) and $Y_2$ (resistance index) data. With the exception of Sequence 1, the fraction of LH realizations selected by using the M indicator is equal/smaller than the fraction selected by using $Y_1$ and $Y_2$.

## Acknowledgements

Dr Fulvio Ciriaco, University of Bari, Italy, is cordially thanked for useful discussions on the computational procedures.

This research has been partially supported by the MIUR (Ministero Italiano Università e Ricerca) program FFABR 2017.

## Notes and references

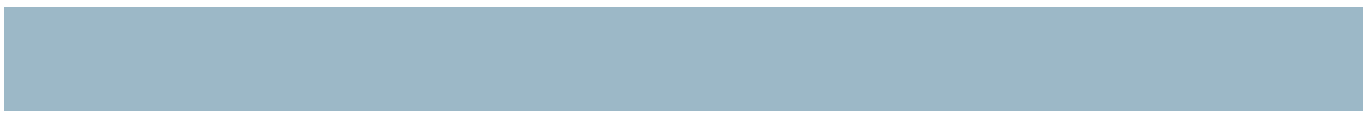